\documentclass[10pt,a4paper,twoside]{article}

\usepackage{graphicx}
\usepackage{amssymb,amsmath,multirow, graphics,color,epstopdf,verbatim}
\usepackage{authblk}
\usepackage{caption}
\usepackage{subcaption}

\newtheorem{theorem}{Theorem}[section]

\newtheorem{example}[theorem]{Example}

\newtheorem{lemma}[theorem]{Lemma}

\begin{document}

\title {New characterization based symmetry tests}

\author[a,b]{V. Bo\v zin\thanks{bozin@matf.bg.ac.rs}}
\affil[a]{\small Faculty of Mathematics, University of Belgrade, Studenski trg 16, Belgrade, Serbia}
\author[a]{B. Milo\v{s}evi\'c\thanks{bojana@matf.bg.ac.rs}}
\author[c,d]{Ya. Yu. Nikitin\thanks{yanikit47@mail.ru}}
\author[a]{M. Obradovi\' c\thanks{marcone@matf.bg.ac.rs}}
\affil[b]{\small Mathematical Institute SANU, Kneza Mihaila 36, Belgrade, Serbia}
\affil[c]{\small Department of Mathematics and Mechanics, Saint-Petersburg State University, Universitetskaia nab. 7/9,
	Saint-Petersburg 199034, Russia}
\affil[d]{\small National Research University - Higher School of Economics, Souza
	Pechatnikov, 16, St.Peters\-burg 190008, Russia}

\date{}

\maketitle



\begin{abstract}
	Two new symmetry tests, of integral and  Kolmogorov type, based on the characterization by squares of linear statistics are proposed.
	The test statistics are related to the family of degenerate U-statistics. Their asymptotic properties are explored.  The maximal eigenvalue,
	needed for the derivation of their logarithmic tail behavior, was calculated  or approximated using techniques from the theory of linear operators
	and the perturbation theory. 
	The quality of the tests is assessed using  the approximate Bahadur efficiency  as well as the simulated powers. The tests are shown to
	be comparable with some recent and classical  tests of symmetry.
\end{abstract}

{\small \textbf{ keywords:} symmetry tests, characterization, degenerate U-statistics, second order Gaussian chaos process,
	approximation of maximal eigenvalue, asymptotic efficiency
	
	}



\section{Introduction}
Consider the classical problem of testing the univariate symmetry with respect to zero. Let $F$ be the distribution function (d.f.)
of an i.i.d. sample $X_1,...,X_n$, and suppose it is continuous. We are interested in testing the hypothesis
\begin{equation}
	\label{symm}
	H_0: 1 - F(x) - F(-x) = 0, \, \, \forall x \in \mathbb{R},
\end{equation}
against the alternative $H_1$ under which the equality in \eqref{symm} is violated at least in one point.

Well-known and simple test statistics for this problem are the sign statistic
and the Wilcoxon signed rank statistic. Their properties are thoroughly explored and described in classical literature, as are some more sophisticated
signed rank statistics (see e.g. \cite{hajekSidak}, \cite{huskova}, \cite{lehmann}, \cite{behboodian1980}).

Another class contains symmetry tests  based on the empirical d.f.'s. Many examples, including the Kolmogorov-Smirnov- and $\omega^2$-type tests,
are described in \cite{nikitinknjiga}. This monograph offers an extensive review of various symmetry tests, together with the calculation of their
efficiencies.

In recent times, introducing tests based on characterizations became a popular direction in goodness-of-fit testing.
Such tests are attractive because they employ some
{\it intrinsic properties} of the probability laws related to the characterization, and therefore they can exhibit high efficiency and power.

The first to introduce such symmetry tests were Baringhaus and Henze in \cite{baringhausHenzeSymmetry}.
They  proposed suitable U-empirical 
Kolmogorov-Smirnov-
and $\omega^2$-type tests of symmetry based on their characterization. The calculation of Bahadur efficiencies, for the Kolmogorov-type test,
was then performed in \cite{nikitinBarHen}, see also \cite{nikitinLDKS}. An integral-type symmetry test, based on the same characterization,
was proposed and analyzed by Litvinova in \cite{litvinovaSimetrija}. 

Recently, Nikitin and Ahsanullah \cite{nikitinAhsanullah} built new tests of symmetry with respect to zero, based on the characterization by Ahsanullah
\cite{ahsanullah}.
This characterization was gene\-ra\-lized and used for construction of similar symmetry tests by Milo\v{s}evi\'c and Obradovi\'c 
\cite{simetrijaSPL}. The quality of all these tests was examined using the Bahadur efficiency, which is applicable to the case of non-normal limiting
distributions.

Here we consider the characterization obtained independently by Wesolowski \cite[Corollary 1]{wesolowski}, and Donati-Martin, Song and Yor
\cite[Lemma 1]{donatimartin}. They proved the following proposition:

{\it Let $X$ and $Y$ be i.i.d. random variables such that $(X-Y)^2$ and $(X+Y)^2$ are equidistributed. Then $X$ and $Y$ are symmetric with respect to zero.}

Our aim is to build the integral- and the Kolmogorov-type U-empirical tests of symmetry based on this characterization; 
to explore their asymptotic properties; and to assess their quality via the approximate
Bahadur efficiency and the simulated powers.

Our test statistics are based on U-statistics. This large class of statistics, which was first defined in the
middle of last century in the problems of the unbiased estimation \cite{halmos}, and its theory established in the seminal paper of Hoeffding \cite{hoeffding},
are very important since numerous well-known statistics belong to this class. The most complete treatment of the theory can be found in
\cite{korolyuk} and \cite{lee}.
In our paper, unlike in many others in this domain of research, the emerging
U-statistics and the families of U-statistics turn out to be {\it degenerate}.
This feature highly complicates the problem and makes it new
and attractive.

The limiting distribution of the underlying U-statistics is the second order Gaussian chaos  (see \cite[Chapter 3]{ledouxTalagrand}). Their tail behavior 
depends on the maximal eigenvalue of the corresponding integral operator. In the case of the uniform distribution we obtain it theoretically. 
When studying the U-em\-pi\-ri\-cal Kolmogorov-type  test, we are forced to work with the family
of degenerate U-statistics depending on a real parameter $t$. Here a challenging problem lies in finding the supremum of the first eigenvalues 
(also depending on $t$) of the corresponding family of integral operators. This is mathematically the most interesting and original point. In
the case of the uniform distribution, we solve it using  a suitable decomposition of the family of operators to some simpler "triangular" operators 
(see Appendix). In the case of other distributions, we approximate the corresponding maximal eigenvalue using the appropriate sequences of discrete linear 
operators.  The application of these mathematical means,
for the first time in this field, is the most innovative and important feature of the paper.


The rest of the paper is organized as follows. In Section 2 we propose the test statistics and study their limiting distributions. Section 3 is 
devoted to the calculation of the approximate Bahadur efficiencies of our tests. In Section 4 we assess the powers of our tests through a 
simulation study. For convenience, some proofs are given in the Appendix.

\section{Test statistics}
Let $F_n^{*}$ be the empirical d.f. of  $|X_1|,...,|X_n|$, and
let $F^{*}(t) = P\{ |X_1| < t \}$ be its theoretical counterpart.
In view of the characterization, we consider the following two test statistics:
\begin{align}
	\bar{J}_n&=\int_{-\infty}^{\infty}(G_n(t)-H_n(t))dF_n^{\ast}(t),\\
	K_n&=\sup_{t} |G_n(t)-H_n(t)|
\end{align}
where
\begin{align*}
	G_n(t)&=\binom n2^{-1}\sum_{i<j}{\rm I}\{|X_i-X_j|<t\},\\
	H_n(t)&=\binom n2^{-1}\sum_{i<j}{\rm I}\{|X_i+X_j|<t\}
\end{align*}
are U-empirical d.f.'s.
After the integration we obtain
\begin{equation*}
	\bar{J}_n=n^{-1}\binom n2^{-1}\sum_{i<j}\sum_{k} {\rm I}\{|X_i-X_j|<|X_k|\}-{\rm I}\{|X_i+X_j|<|X_k|\}.
\end{equation*}

The statistic $\bar{J}_n$ is a hybrid of a U- and a V-statistics. Instead of it, we propose the corresponding U-statistic
$$
J_n = \binom n3^{-1}\sum_{1\le i <j <k \le n} \Phi(X_i,X_j,X_k),
$$
with symmetrized kernel \begin{equation}\label{Phi}
	\begin{aligned}
		\Phi(X_1,X_2,X_3)&=\frac{1}{3!}\sum_{\pi\in\Pi(3)}{\rm I}\{|X_{\pi(1)}-X_{\pi(2)}|<
		|X_{\pi(3)}|\}\\&-{\rm I}\{|X_{\pi(1)}+X_{\pi(2)}|<|X_{\pi(3)}|\},
	\end{aligned}
\end{equation}
where $\Pi(m)$ is the set of all permutations of the set
$\{1,2,\ldots,m\}$.

This statistic is more natural, and, moreover, an unbiased estimator of
$$
\int_{-\infty}^{\infty}\big(P\{|X-Y|<t\}-P\{|X+Y|<t\}\big) dF^{\ast}(t).
$$
Magnified by the factor $n$, $J_n$ and $\bar{J}_n$
have somewhat different limiting distributions. However, in terms of the logarithmic tail behaviour they are equivalent,
and both statistics lead to consistent tests for our hypothesis. 

We consider large values of $|J_n|$ and $K_n$ to be significant.

\subsection{Statistic $J_n$}
The statistic $J_n$ is a U-statistic  with symmetric kernel $\Phi$ given in \eqref{Phi}.
Its first projection on $X_1$ under $H_0$  is equal
to zero, while the second projection on $(X_1,X_2)$, at the point $(s,t)$, is
\begin{equation}\label{projekcija}
	\varphi_F(s,t)=2/3\Big(F\big(|s+t|\big)-F\big(|s-t|\big)\Big),
\end{equation}
where $F$ is the d.f.  of a null symmetric distribution.

\begin{theorem}
	Under $H_0$, the following convergence in distribution holds
	\begin{equation}\label{raspodelaintegralne}
		nJ_n\overset{d}{\to}\binom{3}{2}\sum_{i=1}^{\infty}\nu_{i}(F)\Big(W^2_i-1\Big),
	\end{equation}
	where $\{W_i\}$ is the sequence of i.i.d. standard normal random variables, and $\{\nu_i(F)\}$ is the non-increasing sequence 
	of eigenvalues of the integral operator $\mathcal{J}_F$ with kernel $\varphi_F$.
	
\end{theorem}
\textbf{Proof.} Since the kernel $\Phi$ is bounded and degenerate,
the result follows from the theorem for the asymptotic distribution of U-statistics with
degenerate kernels \cite[Corollary 4.4.2]{korolyuk}. \hfill{$\Box$}

Since our test statistic is not distribution free, the eigenvalues need to be derived for each null distribution. 
In the following theorem we consider the case of
the uniform distribution. 

\begin{theorem}\label{spektarIntegralne}
	Let $F$ be the d.f. of the uniform $U[-\vartheta,\vartheta]$ distribution. Then the sequence of eigenvalues ${\nu_i}$ from 
	\eqref{raspodelaintegralne} are the solutions of the following equation
	
	\begin{equation}\label{eigenInt}
		\frac{\tan\big(\frac{1}{2\sqrt{\nu}}\big)}{6\sqrt{\nu}}-\frac{\cot\big(\frac{1}{2\sqrt{3\nu}}\big)}{2\sqrt{3\nu}}=0.
	\end{equation}
	
\end{theorem}
\textbf{Proof.} 
It is easy to see that our statistic is scale free. Therefore we may suppose that $F$ is the d.f. of the uniform $U[-1,1]$ distribution.

From \eqref{projekcija} it is obvious that the function $\varphi$ is odd, as a function of $s$ as well as a function of $t$. 
Hence, it suffices to present the kernel for $s>0$ and $t>0$.
\begin{equation*}
	\varphi(s,t)=\left\{
	\begin{array}{ll}
		-\frac{2}{3}s, & 0<s<t<1,s+t<1; \\
		\frac{1}{3}(1+s-t), & 0<s<t<1,s+t>1; \\
		\frac{2}{3}t, & 0<t<s<1,s+t<1; \\
		\frac{1}{3}(1-s+t), & 0<t<s<1,s+t>1. \\
	\end{array}
	\right.
\end{equation*}

By definition, the eigenvalues and their eigenfunctions $e$ satisfy
\begin{equation}\label{jednacina}
	\nu e(s)=\mathcal{J}[e(t)]=\int_{-1}^{1}\frac{1}{2}e(t)\varphi(s,t)dt.
\end{equation}
Since the kernel is an odd function, the eigenfunctions must be odd, too. Therefore they can be represented using their Fourier expansion
\begin{equation}\label{furijeova}
	e(t)=\sum_{k=1}^{\infty}a_k u_k(t),
\end{equation}
where $u_k(s)=\sin k\pi s$.

Applying the operator $\mathcal{J}$ to the function $e$ given in \eqref{furijeova}, we have that \eqref{jednacina} is equivalent to
\begin{equation}\label{primenaOperatora}
	\mathcal{J}\Big[\sum_{k=1}^{\infty}a_ku_k(t)\Big]=\nu\sum_{l=1}^{\infty}u_l(s)a_l.
\end{equation}
The left hand side is a function from $L^2[-1,1]$, and its Fourier expansion is
\begin{equation}\label{problematicno}
	\mathcal{J}\Big[\sum_{k=1}^{\infty}a_ku_k(t)\Big]=\sum_{l=1}^{\infty}b_lu_l(s),
\end{equation}
where $b_l=\sum_{k=1}^{\infty}a_k\langle\mathcal{J}[u_k(t)],u_{l}(t)\rangle/||u_{l}(t)||=2\sum_{k=1}^{\infty}\langle\mathcal{J}[u_k(t)],u_{l}(t)\rangle$, 
and $\langle \cdot,\cdot\rangle$ is the scalar product.
After some calculations we get
\begin{equation}\label{koeficijenti}
	\langle \mathcal{J}[u_k],u_l\rangle=\left\{
	\begin{array}{rl}
		\frac{(-1)^{k+l}}{3kl\pi^2}, & k\neq l; \\
		\frac{2}{3k^2\pi^2}-\frac{(-1)^{k}}{6k^2\pi^2}, & k=l.
	\end{array}
	\right.
\end{equation}
From \eqref{primenaOperatora} and \eqref{problematicno} we obtain the system
\begin{equation}\label{koef2}
	2\sum_{k=1}^{\infty}a_k\langle \mathcal{J}[u_k],u_l\rangle=\nu a_l,
\end{equation}
and substituting \eqref{koeficijenti} in \eqref{koef2} we get 
\begin{equation}\label{eqal}
	\nu a_l=2\sum_{k=1}^{\infty}a_k\frac{(-1)^{k+l}}{3\pi^2kl}+\Big(\frac{2}{3\pi^2l^2}-\frac{(-1)^l}{3l^2\pi^2}\Big)a_l=\frac{2(-1)^l}{3\pi^2l}C+\frac{2-(-1)^l}{3l^2\pi^2}a_l,
\end{equation}
where
\begin{equation*}
	C=\sum_{k=1}^{\infty}\frac{(-1)^k}{k}a_k.
\end{equation*}
Transforming \eqref{eqal} we obtain
\begin{equation}\label{koef3}
	\frac{(-1)^l}{l}a_l=\frac{2}{3\pi^2l^2\nu-2+(-1)^l}C.
\end{equation}
Summing both sides of \eqref{koef3} for $l=1,2...$ we obtain
\begin{equation*}
	C=C\sum_{l=1}^{\infty}\frac{2}{3\pi^2l^2\nu-2+(-1)^l}=C\Big(\frac{\tan\big(\frac{1}{2\sqrt{\nu}}\big)}{6\sqrt{\nu}}-\frac{\cot\big(\frac{1}{2\sqrt{3\nu}}\big)}{2\sqrt{3\nu}}+1\Big),
\end{equation*}
from where follows \eqref{eigenInt}.\hfill$\Box$

\subsection{Statistic $K_n$}

For a fixed $t>0$, the expression $K_n^\ast(t)=G_n(t)-H_n(t)$ is a U-statistic with the symmetric kernel

\begin{equation*}
	\Xi(X_1,X_2;t)=I\{|X_1-X_2|<t\}-I\{|X_1+X_2|<t\}.
\end{equation*}

It is easy to see that the kernel is degenerate with the second projection

\begin{equation*}
	\xi(s_1,s_2;t)=I\{|s_1-s_2|<t\}-I\{|s_1+s_2|<t\}.
\end{equation*}


For studying the asymptotics of the statistic $K^{\ast}_n$, it is of interest to consider the integral operator with the same kernel.

For any function  $v\in L^2(\mathbb{R})$ we define it as
\begin{equation}\label{integralniOperator}
	\mathcal{Q}_F(t)[v(x)]=\int_{\mathbb{R}}\xi(x,y;t)v(y)\cdot dF(y).
\end{equation}

Let $\{\nu_i(t;F)\}$ be the sequence of the eigenvalues of the integral operator $\mathcal{Q}_F(t)$.
In the following theorem we give the limiting process of $nK^{\ast}(t)$. This process is called the second order Gaussian chaos process
(see e.g. \cite[Chapter 3]{ledouxTalagrand}).

\begin{theorem}
	Under $H_0$, the limiting process of $nK_n^{\ast}(t)$, $n\to\infty$, is
	\begin{equation}\label{limitingKapa}
		\zeta_F(t)=\sum_{i}\langle\mathcal{Q}_F(t)[e_i(x)],e_i(x) \rangle (W_i^2-1) + \sum_{i\neq j}\langle\mathcal{Q}_F(t)[e_i(x)],e_j(x) \rangle W_iW_j,
	\end{equation}
	where $\{e_i\}$ is an orthonormal basis of $L^2(\mathbb{R})$, and $\{W_i\}$ are i.i.d. standard normal random variables.
\end{theorem}
\textbf{Proof.} Our class of kernels $\xi(x,y;t)$ is Euclidean in the sense of \cite{nolanPollard87}, so the conditions of \cite[Theorem 7]{nolanPollard88}
are satisfied, and \eqref{limitingKapa} follows. \hfill$\Box$

Hence, $nK_n$ converges to the random variable $\sup_{t\in[0,\infty]}|\zeta_F(t)|$.

\section{Approximate Bahadur efficiency}

Let  $\mathcal{G}=\{G(x;\theta)\}$ be the family of d.f.'s with densities $g(x;\theta)$, such that $G(x,\theta)$ is symmetric only for $\theta=0$.
We assume that the d.f.'s from the class $\mathcal{G}$ satisfy the regularity conditions from \cite[Assumptions WD]{nikitinMetron}. 
Denote $h(x)=g'_{\theta}(x;0)$.

Suppose that   $T_n=T_n(X_1,...,X_n)$ is a sequence of   test statistics whose large values are significant, i.e. the null hypothesis $H_0:\theta\in\Theta_0$ 
is rejected in favour of $H_1:\theta\in\Theta_1$, whenever $T_n>t_n$. Let the sequence of  d.f.'s of the test statistic $T_n$ converge in distribution 
to a non-degenerate d.f. $F$. Additionally, suppose that

$$\log(1-F(t))=-\frac{a_Tt^2}{2}(1+o(1)),\;\;t\to \infty,$$
and the limit in probability under the alternative
$$
\lim_{n\rightarrow\infty}T_n/\sqrt{n}=b_T(\theta)>0
$$ exists  for $\theta \in \Theta_1$.

The approximate relative  Bahadur efficiency with respect to another test statistic $V_n=V_n(X_1,...,X_n)$ is defined as
\begin{equation*}
	e^{\ast}_{T,V}(\theta)=\frac{c^{\ast}_T (\theta)}{c^{\ast}_V (\theta)},
\end{equation*}
where
\begin{equation}\label{BASlope}
	c^{\ast}_T(\theta)=a_Tb_T^2(\theta)
\end{equation} is called the Bahadur  approximate slope of $T_n$. This is a popular measure of the test efficiency
suggested by Bahadur in \cite{bahadur1960}.

\subsection{Integral-type test}

In the case of our integral-type test statistic, the role of $T_n$ is played by the statistic $\widetilde{J}_n=\sqrt{n|J_n|}$.
Its Bahadur approximate slope is obtained in the following lemma.

\begin{lemma} For the statistic $\widetilde{J}_n$ and a given alternative density $g(x;\theta)$ from $\mathcal {G}$, the Bahadur approximate slope
	satisfies the relation
	\begin{equation*}
		c^{\ast}_{\widetilde{J}}(\theta)  \sim \frac{b_{J}(\theta)}{3\nu_1},
	\end{equation*}
	where $b_J(\theta)$ is the limit in ${P_\theta}$ probability of $J_n$, 
	and $\nu_1$ is the largest eigenvalue of the sequence $\{\nu_i(F)\}$ in \eqref{raspodelaintegralne}.
\end{lemma}
\textbf{Proof.}
Using the result of Zolotarev \cite{zolotarev}, we have that the logarithmic tail behavior of $\widetilde{J}$ is
\begin{equation*}
	\log(1-F_{\widetilde{J}}(x))=-\frac{x^2}{6\nu_1}+o(x^2),\;\; x\to \infty,
\end{equation*}
and hence, $\widetilde{a}_{\widetilde{J}}=\frac{1}{3\nu_1}$.
The limit in probability of $\widetilde{J}_n/\sqrt{n}$ is
\begin{equation*}
	\widetilde{b}_{\widetilde{J}}(\theta)=|b_J(\theta)|^{\frac{1}{2}}.
\end{equation*}
Inserting the expressions for $\widetilde{a}_{\widetilde{J}}$ and
$\widetilde{b}_{\widetilde{J}}(\theta)$ into \eqref{BASlope}, we obtain the statement of the lemma.\hfill{$\Box$}

The largest eigenvalue in the case of the uniform distribution is calculated from  \eqref{eigenInt}, and is equal to 0.1898. 
The equations for the eigenvalues of the operator

\begin{equation}\label{operatorJ}\mathcal{J}_F[v(x)]=\int_{R}\varphi_F(x,y)v(y)\cdot dF(y)\end{equation}
for other distributions $F$ are too complicated to derive. Thus we calculate the largest eigenvalues using the following approximation.
First, notice that the "symmetrized" operator 
\begin{equation}\label{operatorS}
	\begin{aligned}\mathcal{S}_F[v(x)]&=\sqrt{F'(x)}\mathcal{J}_F\Big(\frac{v(x)}{F'(x)}\Big)\\&=\int_{R}\varphi_F(x,y)v(y)\sqrt{F'(x)}\sqrt{F'(y)}dy.\end{aligned}
\end{equation}
has the same spectrum as the operator $\mathcal{J}_F$.

Consider the case where  $\inf_{x}\{x: F(x)=1\}=A<\infty$.
The sequence of symmetric linear operators defined by $(2m+1)\times(2m+1)$ matrices $M^{(m)}_F=||m_{i,j}^{(m)}||,\; |i|\leq m,|j|\leq m$, where
\begin{equation*}
	m_{i,j}^{(m)}=\varphi_F\Big(\frac{Ai}{m},\frac{Aj}{m}\Big)\sqrt{\Big(F\Big(\frac{A(i+1)}{m}\Big)-F\Big(\frac{Ai}{m}\Big)\Big)}\sqrt{\Big(F\Big(\frac{A(j+1)}{m}\Big)-F\Big(\frac{Aj}{m}\Big)\Big)},
\end{equation*}
converges in norm to $\mathcal{S}_F$.

Indeed, for a function $v\in L^2[-A,A]$, the operator $M^{(m)}_F$, for $x\in[Ai/m,A(i+1)/m)$,
can be written as
\begin{align*}
	M^{(m)}_F[v](x)=\sum_{j=-m}^{m}\Bigg(\varphi_F\Big(\frac{Ai}{m},\frac{Aj}{m}\Big)\sqrt{\Big(F\Big(\frac{A(i+1)}{m}\Big)-F\Big(\frac{Ai}{m}\Big)\Big)}\\
	\cdot\sqrt{\Big(F\Big(\frac{A(j+1)}{m}\Big)-F\Big(\frac{Aj}{m}\Big)\Big)}v\Big(\frac{Aj}{m}\Big)\Bigg).
\end{align*}
This sum converges to the Lebesgue integral from \eqref{operatorS}. Since this is true for every function $v$, then
\begin{equation*}
	\|\mathcal{S}_F-M^{(m)}_F\|=\sup\limits_{\|v\|=1}\|\mathcal{S}[v]-M^{(m)}_F[v]\|\to 0, m\to\infty.
\end{equation*}

The operators $\mathcal{S}$ and $M^{(m)}_F$ are symmetric and self-adjoint, and the norm of their difference tends to zero as $m$ tends to infinity. 
Using the perturbation theory, see \cite[Theorem 4.10, page 291]{kato}, we have that the spectra of these two operators are at the distance that 
tends to zero. Hence, $\nu_1^{(m)}(F)$ -- the sequence of the largest eigenvalues of $M^{(m)}_F$ -- must converge to $\nu_1(F)$, the largest eigenvalue of 
$\mathcal{S}$.

In the case where $\inf_{x}\{x: F(x)=1\}=\infty$, we consider its truncation to the interval $[-A,A]$, such that $F(A)>1-\varepsilon$ for a desired 
value of $\varepsilon>0$. The values for some common symmetric distributions are given in Table \ref{fig:lambdeInt}.

\begin{table}[htbp]
	\centering
	\bigskip
	\caption{Approximative values of the largest eigenvalues of the operator $\mathcal{J}_F$}
	\bigskip
	\centering
	\begin{tabular}{ccccc}
		\hline\noalign{\smallskip}
		$F$ & $A$ & $\nu_1^{(1000)}(F)$\\\hline
		Normal & 10 & 0.138\\
		Logistic & 30 & 0.126\\
		Laplace & 30 & 0.101\\\hline
	\end{tabular}
	\label{fig:lambdeInt}
\end{table}

\medskip

Concerning the limit in probability $b_J(\theta)$, we give its formula in the following lemma for  
the alternatives close to the uniform distribution. Its proof follows from \cite{nikitinMetron}.

\begin{lemma}\label{bJn} Under a close alternative $g(x;\theta)$ from $\mathcal {G}$, such that $G(x;0)=F(x)$, 
	the limit in probability of $J_n$ is
	\begin{equation*}
		b_J(\theta)=3\int\limits_{\mathbb{R}^2}\varphi_F(x,y)h(x)h(y)dxdy\cdot\theta^2+o(\theta^2),
	\end{equation*}
	as $\theta\to 0$, where $h(x)=g_\theta'(x;0)$.
\end{lemma}







We consider null d.f.'s to be uniform, normal, logistic and Laplace; and the following  alternative distributions close to a null d.f. $F$:

\begin{itemize}
	\item a Lehmann alternative with d.f.
	\begin{equation}\label{leman}
		G_1(x;\theta)=F^{1+\theta}(x), \;\; \theta>0;
	\end{equation}
	\item a first Ley-Paindaveine \cite{ley_paindaveine_2008} alternative  with d.f.
	\begin{equation}\label{lp1}
		G_2(x;\theta)=F(x)e^{-\theta(1-F(x))}, \;\; \theta>0;
	\end{equation}
	\item a second Ley-Paindaveine alternative \cite{ley_paindaveine_2008} with d.f.
	\begin{equation}\label{lp2}
		G_3(x;\theta)=F(x)-\theta\sin\big(\pi F(x)\big), \;\; \theta\in[0,\pi^{-1}];
	\end{equation}
	\item a contamination (with $G_1$) alternative  with  d.f.
	\begin{equation}\label{contLeman}
		G_4(x;\theta,\beta)=(1-\theta)F(x)+\theta F^{\beta}(x),\;\;\beta>1,\;\theta\in[0,1];
	\end{equation}
	\item a location alternative with d.f.
	\begin{equation}\label{location}
		G_5(x;\theta)=F(x-\theta);
	\end{equation}
	
	\item a contamination (with shift) alternative  with  d.f.
	\begin{equation}\label{contLoc}
		G_6(x;\theta,\beta)=(1-\theta)F(x)+\theta F(x-\beta),\;\;\;\beta>0,\;\theta\in[0,1];
	\end{equation}
	\item a skew alternative in the sense of Azzalini \cite{azzalini} with density 
	\begin{equation}\label{azalini}
		g_7(x;\theta)=2F(\theta x)f(x).
	\end{equation}
	
\end{itemize}

\begin{example}
	Let the alternative distribution be  $G_1(x;\theta)$ when $F$ is the d.f. of the uniform distribution. In this case we have
	
	\begin{align*}
		h(x)=\frac{1}{2}+\frac{1}{2}\log\Big(\frac{x+1}{2}\Big).
	\end{align*}
	By Lemma \ref{bJn}, after calculating the corresponding integral,
	we have that $b_J(\theta)\sim 0.348\theta^2$, $\theta\to0$. The largest solution of \eqref{eigenInt}  is $\nu_1\approx 0.1898$. 
	Therefore the Bahadur approximate slope is
	
	\begin{align*}
		c^{\ast}_{\widetilde{J}}(\theta)\sim\frac{0.348}{3\nu_1}\cdot \theta^2\approx 0.611\cdot \theta^2,\theta\to 0.
	\end{align*}
\end{example}

The calculations are similar for the other alternatives. The Bahadur approximate indices (the leading coefficient in the Maclaurin expansion of the Bahadur 
approximate slope)  are presented in Table \ref{fig:eff} at the end of this section, together with the results for the Kolmogorov-type test.

\subsection{ Kolmogorov-type test}

Similarly to the integral-type  statistic, we study  tail behavior of the limiting distribution of the statistic $\widetilde{K}_n=\sqrt{nK_n}$. 
Notice that $\widetilde{K}_n$ converges to $\sup_{t}|\zeta_F(t)|^{1/2}$.
\begin{theorem}\label{behaviourKn}
	For the limiting distribution of the test statistic $\widetilde{K}_n$,  it holds true that
	\begin{equation}\label{korenKapa}
		\log(1-F_{\widetilde{K}}(x)) \sim -\frac{x^2}{2\varkappa_0(F)},\;x\to \infty,
	\end{equation}
	where $\varkappa_0(F)=\sup_{t}|\nu_1(t;F)|$, and $\{\nu_1(t;F)\}$ is the set of  eigenvalues of the family of operators
	$\mathcal{Q}_F(t),\;t>0$, with the largest absolute value.
\end{theorem}
\textbf{Proof.}
The limiting process of $nK_n^{\ast}$ is the second order Gaussian chaos process $|\zeta_F(t)|$. The tail behaviour of its supremum $nK_n$ 
is obtained in \cite[Corollary 3.9]{ledouxTalagrand}. The constant $\sigma$ appearing there is equal to the supremum of the maximal eigenvalues of the
corresponding linear operator. This is because
$$
\sup_{||e||\leq1} \sup_{t}\langle\mathcal{Q}_F(t)[e(x)],e(x)\rangle=\sup_{t }\sup_{||e||\leq1}\langle\mathcal{Q}_F(t)[e(x)],e(x)\rangle = \sup_{t}\nu_1(t;F).
$$
Writing it in our notation, we obtain

\begin{equation}
	\lim_{x\to\infty}\frac{1}{x}\log P\{\sup_{t}|\zeta_F(t)|>x\}=-\frac{1}{2\varkappa_0(F)},
\end{equation}
and transforming the variable we get \eqref{korenKapa}.
\hfill{$\Box$}


The following lemma gives us the limit in probability of the statistic $K_n$.

\begin{lemma}\label{bKn} Under a close alternative $g(x;\theta)$ from $\mathcal {G}$, such that $G(x;0)=F(x)$, 
	the limit in probability of $K_n$ is
	\begin{equation*}
		b(\theta)=\sup_{t>0}\bigg|\underset{\mathbb{R}^2}{\int\!\!\!\int}\xi(s_1,s_2;t)h(s_1)h(s_2)ds_1ds_2\bigg|\cdot\theta^2+o(\theta^2), \, \theta\to 0,
	\end{equation*}
\end{lemma}
\textbf{Proof}. Denote by $a(t;\theta)$ the limit in probability of the statistic $K_n^\ast(t)$ under the alternative $g(x;\theta)$.
Using the Glivenko--Cantelli theorem for U-statistics \cite{helmersjanssen}, we have that $b(\theta)=\sup_{t>0}a(t;\theta)$, where

\begin{equation*}
	a(t;\theta)=\underset{\mathbb{R}^2}{\int\!\!\!\int}\Xi(x,y;t)g(x;\theta)g(y;\theta)dxdy.
\end{equation*}
It is easy to show that $a(t;0)=a'(t;0)=0$. The second derivative of $a(t;\theta)$ along $\theta$ at $\theta=0$ is

\begin{equation*}
	a''(t;0)=2\underset{\mathbb{R}^2}{\int\!\!\!\int}\xi(x,y;t)h(x)h(y)dxdy.
\end{equation*}
Expanding $a(t;\theta)$ in the Maclaurin series completes the proof. \hfill$\Box$.

Finally, the Bahadur approximate slope is given in the following theorem.

\begin{theorem} For the statistic $\widetilde{K}_n$ and a given alternative density $g(x;\theta)$ from $\mathcal {G}$, the Bahadur approximate slope
	satisfies the relation 
	\begin{equation*}
		c^{\ast}_{\widetilde{K}}(\theta) = \frac{1}{\varkappa_0}
		\sup_{t>0}\bigg|\underset{\mathbb{R}^2}{\int\!\!\!\int}\xi(s_1,s_2;t)h(s_1)h(s_2)ds_1ds_2\bigg|\cdot\theta^2+o(\theta^2), \, \theta\to 0,
	\end{equation*}
	where $\varkappa_0$ is given in \eqref{kapa0}.
\end{theorem}
\textbf{Proof}. Using Theorem \ref{behaviourKn} and Lemma \ref{bKn}, and the same arguments as in the case of the statistic $\widetilde{J}_n$, we get
the statement of the theorem. \hfill$\Box$.

\begin{figure}[h!]
	\begin{center}
		\includegraphics[scale=0.6]{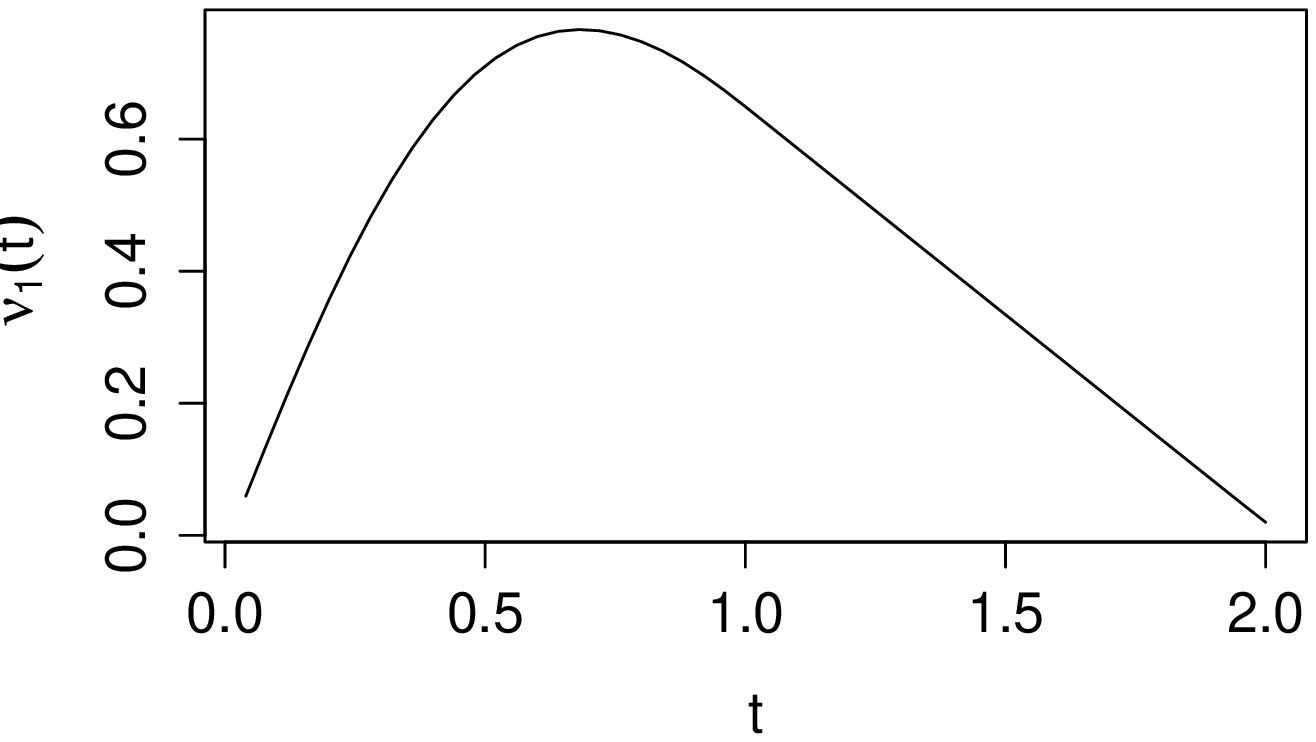}\caption{Maximal eigenvalue function $\nu_1^{(1000)}(t)$  }
		\label{fig: maxEigen}
	\end{center}
\end{figure}

We apply the same approximation procedure  used in the case of the operator $\mathcal{J}_F$. 
For the eigenvalues  of the operator $\mathcal{Q}_F(t)$,  for $n=1000$, we get the functions $\nu_1^{(1000)}(F;t)$ in the case of uniform, 
normal, logistic and Laplace distributions. Since in all the cases the functions have the same shape, we show only the function
$\nu_1^{(1000)}(F;t)$ for the uniform distribution case
(Figure \ref{fig: maxEigen}).

From Figure \ref{fig: maxEigen} we can see that the function has a unique maximum $\varkappa_0\approx 0.766$ for $t\approx 2/3$. 
In this case, however, we are able to  derive $\varkappa_0$ theoretically in the following  lemma.


\begin{lemma}\label{lemaBozin} Let $F(x)=(x+1)/2$, $x\in [-1,1]$ and let $\mathcal{Q}(t)$ be the integral operator 
	\eqref{integralniOperator} corresponding to this $F$.
	Let  $\varkappa_0=\sup_{t\in[0,2]}|\nu_1(t)|$, where $\{\nu_1(t), 0 \leq t \leq 2\} $ is the set of eigenvalues of the family of operators
	$\mathcal{Q}(t),\;t\in(0,2)$
	with the largest absolute value. Then
	\begin{equation}\label{kapa0}
		\varkappa_0=\nu_1\Big(\frac{2}{3}\Big)=\frac{\sqrt{2}}{3}\Big(\arctan\frac{1}{\sqrt{2}}\Big)^{-1}.
	\end{equation}
\end{lemma}
The proof is given in the Appendix.

For the other distributions, we rely on the values obtained using the approximation. We present the obtained values in  Table \ref{fig:lambdeKS}.

\begin{table}[htbp]
	\centering
	\bigskip
	\caption{Approximative values of the largest eigenvalues of the operator $\mathcal{Q}_F(t)$}
	\bigskip
	\centering
	\begin{tabular}{ccccc}
		\hline\noalign{\smallskip}
		$F$ & $A$ & $\sup_{t}\nu_1^{(1000)}(F;t)$\\\hline
		Normal & 10 & 0.629\\
		Logistic & 30 & 0.596\\
		Laplace & 30 & 0.516 \\\hline
	\end{tabular}
	\label{fig:lambdeKS}
\end{table}

\begin{example}
	Let the alternative distribution be  $G_3(x;\theta)$.
	In this case we have
	
	\begin{align*}
		h(x)=\frac{\pi}{2}\sin\Big(\frac{\pi x}{2}\Big).
	\end{align*}
	By Lemma \ref{bKn}, after calculating the corresponding integral,
	we have that
	\begin{equation*}
		b_J(\theta)\sim \sup_{t>0}\Big|\pi(2-t)\sin\Big(\frac{\pi t}{2}\Big)\Big|\theta^2,\;\;\theta\to 0.
	\end{equation*}
	The supremum is attained at $t=0.708$ and is approximately equal to $3.639$. The Bahadur approximate slope is then
	\begin{align*}
		c^{\ast}_{\widetilde{J}}(\theta)\sim 4.752\cdot\theta^2,\theta\to0.
	\end{align*}
\end{example}

The calculations are similar for the other alternatives.
In Table \ref{fig:eff}, we give the values of the local approximate indices.
Naturally, we confine ourselves to the cases where the alternatives \eqref{leman}-\eqref{azalini}
belong to the class $\mathcal{G}$, i.e. satisfy the regularity conditions.

For comparison purpose we also include Bahadur indices of some competitor tests. We choose  some recent characterization based symmetry tests from  \cite{simetrijaSPL} and \cite{nikitinAhsanullah} (labeled respectively as $c_{\mathrm{MOI}_k}$ and $c_{\mathrm{NAI}_k}$ for integral-type tests, and $c_{\mathrm{MOK}_k}$ and $c_{\mathrm{NAK}_k}$ for Kolmogorov-type tests), as well as  the classical sign test ($c_S$). The relative efficiency of any two tests can be calculated as the ratio of their Bahadur indices.


\begin{table}[!htbp]
	\caption{Local approximate Bahadur indices of test statistics}
	\centering
	\bigskip
	\begin{tabular}{ccccccccccc}
		\hline\noalign{\smallskip}
		null& alter. &$c_{J_n}$&$c_{\mathrm{MOI}_1}$ & $c_{\mathrm{MOI}_2}$& $c_{\mathrm{NAI}_4}$ & $c_{K_n}$& $c_{\mathrm{MOK}_1}$ & $c_{\mathrm{MOK}_2}$& $c_{\mathrm{NAK}_4}$ & $c_S$ \\\hline
		\multirow{4}{1pt}{\rotatebox[origin=c]{90}{Uniform}}	&$g_1$&0.611&0.791&0.756&0.793&0.596&0.615&0.631&0.596&0.480\\
		&$g_2$&0.307&0.312&0.327&0.308&0.303&0.250&0.288&0.232&0.250\\
		&$g_3$&4.788&4.281&4.659&4.183&4.752&3.445&4.211&3.132&4.000\\
		&$g_4(3)$&0.691&0.703&0.736&0.692&0.682&0.569&0.648&0.522&0.563\\\hline
		\multirow{7}{1pt}{\rotatebox[origin=c]{90}{Normal}}&$g_1$&0.545&0.791&0.756&0.793&0.572&0.615&0.631&0.596&0.480\\
		&$g_2$&0.292&0.312&0.327&0.308&0.298&0.250&0.288&0.232&0.250\\
		&$g_3$&4.614&4.281&4.659&4.183&4.610&3.445&4.211&3.132&4.000\\
		&$g_4(3)$&0.657&0.703&0.736&0.692&0.670&0.569&0.648&0.522&0.563\\
		&$g_5$&0.730&0.977&0.957&0.975&0.760&0.764&0.810&0.733&0.630\\
		&$g_6(1)$&0.508&0.930&0.828&0.945&0.571&0.717&0.668&0.721&0.466\\
		&$g_7$&0.461&0.622&0.610&0.621&0.490&0.487&0.516&0.467&0.405\\
		
		\hline
		\multirow{7}{1pt}{\rotatebox[origin=c]{90}{Logistic}}&$g_1$&0.509&0.791&0.756&0.793&0.544&0.615&0.631&0.596&0.480\\
		&$g_2$&0.280&0.312&0.327&0.308&0.284&0.250&0.288&0.232&0.250\\
		&$g_3$&4.471&4.281&4.659&4.183&4.437&3.445&4.211&3.132&4.000\\
		&$g_4(3)$&0.630&0.703&0.736&0.692&0.640&0.569&0.648&0.522&0.563\\
		&$g_5$&0.280&0.312&0.327&0.308&0.284&0.250&0.288&0.232&0.250\\
		&$g_6(1)$&0.240&0.311&0.313&0.309&0.251&0.250&0.270&0.237&0.214\\
		&$g_7$&0.509&0.791&0.756&0.793&0.544&0.615&0.631&0.596&0.480\\
		\hline
		
		\hline
		\multirow{5}{1pt}{\rotatebox[origin=c]{90}{Laplace}}&$g_1$&0.495&0.791&0.756&0.793&0.523&0.615&0.631&0.596&0.480\\
		&$g_2$&0.275&0.312&0.327&0.308&0.273&0.250&0.288&0.232&0.250\\
		&$g_3$&4.391&4.281&4.659&4.183&4.246&3.445&4.211&3.132&4.000\\
		&$g_4(3)$&0.618&0.703&0.736&0.692&0.612&0.569&0.648&0.522&0.563\\
		&$g_6(1)$&0.511&0.584&0.617&0.574&0.529&0.534&0.581&0.492&0.400\\
		
		\hline
	\end{tabular}
	\label{fig:eff}
\end{table}

We find that our tests, in all cases, are more efficient than the sign test. In comparison to the recent characterization based tests, in some cases (e.g. $g_3$ alternative for the uniform distribution) they outperform all the others, while in some other cases (e.g. $g_1$ for normal distribution)  new tests are the least efficient. Moreover, as it is often the case, no test
is uniformly the most efficient.

When comparing two new tests to each other, we can notice that in the case of the uniform null distribution,
the integral-type test is more efficient than the Kolmogorov-type test. This is a widespread situation in the comparison of tests, see \cite{nikitinknjiga}.
However, for the other null distributions, it is mostly the other way around.


\section{Power study}

In Tables \ref{fig:pow20} and \ref{fig:pow50},  we present the empirical sizes and powers of our tests ($J_n$ and $K_n$)  against  the alternatives $g_5$, $g_6(1)$, and $g_7$, for 
some values of  parameter $\theta$.

The null distributions are normal, Laplace, logistic, and Cauchy.  The simulated powers for $J_n$ and $K_n$, at the  level of significance of 0.05,
are  obtained using a warp-speed Monte Carlo bootstrap procedure \cite{warpspeed} given below.

\paragraph{Warp speed Monte Carlo bootstrap algorithm}
\begin{enumerate} 
	\item[(i)] Generate the sequence $x_1,...,x_n$ from  an alternative distribution  and compute  the value of the test statistic $T_n(x_1,...,x_n)$;
	\item[(ii)] Generate $y^*_k=x_ku_k, \ k=1,...,n$, using i.i.d  sequence of   Rademacher random variables $u_k$ taking values -1 and 1 with equal 
	probabilities, to obtain the symmetrized sampling distribution;
	\item[(iii)]  Calculate the value of  the test statistic $T^*_n=T_n(y_1^*,...,y^*_n)$;
	\item[(iv)] Repeat the steps (i)-(iii)  $N$ times, and obtain the empirical sampling distributions of $T_n$  and $T^*_n$ that correspond to the 
	alternative and the null distribution of the test statistic, respectively;
	\item[(v)] Calculate the empirical power as the percentage  of values of $T^*_n$ greater than the 95th percentile of the empirical sampling distribution
	of $T_n$.
\end{enumerate}

The procedure is done for $N=10000$ replications, for the sample sizes of 20 and 50.
For comparison purpose, we also include the same characterization based tests as in Table \ref{fig:eff}, as well as the classical Kolmogorov-Smirnov symmetry test (KS)
and the sign test (S), whose powers are calculated using the standard Monte Carlo procedure with 10000 replications.

From Tables \ref{fig:pow20} and  \ref{fig:pow50}
one can see that the empirical sizes of our tests are satisfactory. Besides, $J_n$ and $K_n$ have almost equal empirical 
powers for all the alternatives. 

We can observe that our tests have the highest powers in the case of the contamination  alternative $g_6(1)$, for all nulls, for smaller sample sizes ($n=20$), and for the logistic null for $n=50$. Similar conclusion can be made for the location alternative $g_5$ of the Cauchy null, for both sample sizes.

In other cases, while not being uniformly the best, the powers of our tests are comparable to all the competitors'.

\begin{table}[!htbp]
	\centering
	\bigskip
	\caption{ Empirical sizes and powers  at 0.05 level of significance, $n=20$ }
	\bigskip
	\centering
	\begin{tabular}{ccccccccccccccccc}
		\hline\noalign{\smallskip}
		null & alter. & $\theta$ & $J_n$ & $K_n$ & \footnotesize MOI$_1$&\footnotesize MOI$_2$ & \footnotesize NAI$_4$ &\footnotesize  MOK$_1$&\footnotesize  MOK$_2$ & \footnotesize NAK$_4$ &  KS & S \\
		\hline

		\multirow{9}{1pt}{\rotatebox[origin=c]{90}{Normal}}&$g_5$& 0 & 0.05 & 0.05 & 0.05 & 0.05 & 0.05 & 0.05 & 0.04 & 0.03 & 0.02 & 0.04  \\
		
		& $g_5$ & 0.25  & 0.13 & 0.13 & 0.22 & 0.21 & 0.19 & 0.07 & 0.09 & 0.11 &0.09 & 0.13  \\
		&  $g_5$ & 0.5& 0.43 & 0.43 & 0.59 & 0.58 & 0.55 & 0.26 & 0.35 & 0.34 & 0.35 & 0.38  \\
		&  $g_5$ & 0.75 & 0.77 & 0.77 & 0.90 & 0.90 & 0.88 & 0.58 & 0.71 & 0.65 & 0.71 & 0.71  \\
		
		& $g_6(1)$ & 0.1  & 0.14 & 0.14 & 0.08 & 0.08 & 0.07 & 0.03 & 0.04 & 0.04 & 0.11 & 0.13  \\
		&  $g_6(1)$ & 0.25& 0.28 & 0.28 & 0.19 & 0.18 & 0.16 & 0.07 & 0.08 & 0.09 & 0.25 & 0.26  \\
		
		& $g_7$ & 0.5  & 0.28 & 0.27 &0.40 & 0.40 & 0.37 & 0.14 & 0.21 & 0.21 & 0.20 & 0.24 \\
		& $g_7$ & 0.75  & 0.50 & 0.49 & 0.67 & 0.64 & 0.63 & 0.31 & 0.41 & 0.41 & 0.40 & 0.44 \\
		& $g_7$ & 1  & 0.68 & 0.69 & 0.85 & 0.84 & 0.82 & 0.50 & 0.62 & 0.61 & 0.61 & 0.62  \\
		
		\hline
		\multirow{9}{1pt}{\rotatebox[origin=c]{90}{Laplace}}&$g_5$& 0 & 0.05 & 0.05 & 0.05 & 0.05 & 0.05 & 0.04 & 0.04 & 0.03  & 0.02 & 0.04  \\
		& $g_5$ & 0.25 & 0.15 & 0.14 & 0.15 & 0.16 & 0.13 & 0.04 & 0.06 & 0.06 & 0.06 & 0.09  \\
		&  $g_5$ & 0.5 & 0.42 & 0.40 & 0.38 & 0.42 & 0.34 & 0.11 & 0.21 & 0.15 & 0.18 & 0.23  \\
		&  $g_5$ & 0.75 & 0.69 & 0.69 & 0.65 & 0.70 & 0.60 & 0.26 & 0.44 & 0.30 & 0.36 & 0.43  \\
		& $g_6(1)$ & 0.1  & 0.14 & 0.13 & 0.07 & 0.07 & 0.06 & 0.03 & 0.04 & 0.04 & 0.10 & 0.12  \\
		&  $g_6(1)$ & 0.25& 0.27 & 0.26 & 0.14 & 0.14 & 0.12 & 0.04 & 0.06 & 0.07 & 0.23 & 0.24  \\
		& $g_7$ & 0.5 & 0.31 & 0.31 & 0.50 & 0.49 & 0.45 & 0.21 & 0.27 & 0.29 & 0.26 & 0.30 \\
		& $g_7$ & 0.75 & 0.51 & 0.51 & 0.71 & 0.72 & 0.67 & 0.36 & 0.47 & 0.46 & 0.44 & 0.47\\
		& $g_7$ & 1 & 0.63 & 0.64 & 0.84 & 0.84 & 0.82 & 0.51 & 0.61 & 0.61 & 0.60 & 0.62  \\
		\hline
		\multirow{9}{1pt}{\rotatebox[origin=c]{90}{Logistic}}&$g_5$& 0 & 0.05 & 0.05 & 0.06 & 0.05 & 0.05 & 0.04 & 0.05 & 0.03 & 0.02 & 0.04  \\
		
		& $g_5$ & 0.25  & 0.09 & 0.08 & 0.10 & 0.11 & 0.10 & 0.03 & 0.05 & 0.05 & 0.04 & 0.07 \\
		&  $g_5$ & 0.5 & 0.20 & 0.20 & 0.26 & 0.26 & 0.23 & 0.08 & 0.12 & 0.12 & 0.13 & 0.16  \\
		&  $g_5$ & 0.75 & 0.36 & 0.35 & 0.46 & 0.48 & 0.43 & 0.17 & 0.26 & 0.23 & 0.28 & 0.34  \\
		
		& $g_6(1)$ & 0.1  & 0.10 & 0.10 & 0.06 & 0.07 & 0.07 & 0.04 & 0.05 & 0.05 & 0.07 & 0.09  \\
		&  $g_6(1)$ & 0.25& 0.27 & 0.26 & 0.10 & 0.11 & 0.10 & 0.04 & 0.05 & 0.06 & 0.14 & 0.17  \\
		
		& $g_7$ & 0.5 & 0.28 & 0.27 & 0.43 & 0.43 & 0.40 & 0.16 & 0.22 & 0.24 & 0.22 & 0.26  \\
		& $g_7$ & 0.75 & 0.48 & 0.48 & 0.69 & 0.68 & 0.65 & 0.34 & 0.43 & 0.43 & 0.41 & 0.45  \\
		& $g_7$ & 1 & 0.66 & 0.66 & 0.84 & 0.85 & 0.83 & 0.51 & 0.61 & 0.60 & 0.61 & 0.61 \\
		
		\hline
		\multirow{9}{1pt}{\rotatebox[origin=c]{90}{Cauchy}}&$g_5$& 0 & 0.06 & 0.06 & 0.05 & 0.05 & 0.05 & 0.04 & 0.04 & 0.03 & 0.02 & 0.04  \\
		
		& $g_5$ & 0.25 & 0.11 & 0.10 & 0.09 & 0.10 & 0.08 & 0.03 & 0.04 & 0.04 & 0.06 & 0.09 \\
		&  $g_5$ & 0.5 & 0.28 & 0.25 & 0.17 & 0.20 & 0.16 & 0.04 & 0.09 & 0.06 & 0.19 & 0.24 \\
		&  $g_5$ & 0.75 & 0.47 & 0.45 & 0.29 & 0.34 & 0.25 & 0.07 & 0.16 & 0.09 & 0.36 & 0.44  \\
		
		& $g_6(1)$ & 0.1  & 0.11 & 0.11 & 0.06 & 0.07 & 0.06 & 0.03 & 0.04 & 0.03 & 0.07 & 0.10  \\
		&  $g_6(1)$ & 0.25& 0.21 & 0.20 & 0.08 & 0.10 & 0.08 & 0.03 & 0.04 & 0.04 & 0.16 & 0.19  \\
		
		& $g_7$ & 0.5 & 0.21 & 0.21 & 0.30 & 0.32 & 0.26 & 0.10 & 0.15 & 0.14 & 0.19 & 0.24  \\
		& $g_7$ & 0.75 & 0.38 & 0.37 & 0.50 & 0.55 & 0.46 & 0.18 & 0.31 & 0.24 & 0.40 & 0.44 \\
		& $g_7$ & 1 & 0.53 & 0.51 & 0.69 & 0.72 & 0.65 & 0.30 & 0.48 & 0.37 & 0.57 & 0.62  \\
		
		\hline

		\hline
		
	\end{tabular}
	\label{fig:pow20}
\end{table}

\begin{table}[!htbp] 
	\centering
	\bigskip
	\caption{ Empirical sizes and powers  at 0.05 level of significance, $n=50$ }
	\bigskip
	\centering
	\begin{tabular}{ccccccccccccccccc}
		\hline\noalign{\smallskip}
		null & alter. & $\theta$ & $J_n$ & $K_n$ & \footnotesize MOI$_1$&\footnotesize MOI$_2$ & \footnotesize NAI$_4$ &\footnotesize  MOK$_1$&\footnotesize  MOK$_2$ & \footnotesize NAK$_4$ &  KS & S \\
		\hline
		\multirow{9}{1pt}{\rotatebox[origin=c]{90}{Normal}}& $g_5$& 0 &  0.05 & 0.05 & 0.05 & 0.05 &0.05 &0.05 &0.05 &0.05 & 0.05 & 0.03 \\
		& $g_5$ & 0.25  & 0.31 & 0.30 & 0.53 & 0.54 & 0.52 & 0.21 & 0.35 & 0.32 & 0.32 & 0.23 \\
		&  $g_5$ & 0.5&  0.84 & 0.85 & 0.96 & 0.96 & 0.96 & 0.76 & 0.88 & 0.85 & 0.86 & 0.74 \\
		&  $g_5$ & 0.75 & 1.00 & 1.00 & 1.00 & 1.00 & 1.00 & 0.99 & 1.00 & 0.99 & 1.00 & 0.98 \\
		
		& $g_6(1)$ & 0.1  & 0.15 & 0.14 &0.11 &0.09 &0.10 &0.06 &0.08 &0.09 & 0.15 & 0.13 \\
		&  $g_6(1)$ & 0.25& 0.29 & 0.29 & 0.36& 0.32& 0.36&0.20 &0.28 &0.30 & 0.29 & 0.28 \\
		
		& $g_7$ & 0.5  & 0.61 & 0.60 & 0.75 & 0.73 & 0.75 & 0.55 & 0.66 & 0.65 & 0.63 & 0.49 \\
		& $g_7$ & 0.75  & 0.90 & 0.90 & 0.96 & 0.96 & 0.96 & 0.87 & 0.93 & 0.92 & 0.91 & 0.80 \\
		& $g_7$ & 1  & 0.98 & 0.98 & 1.00 & 1.00 & 1.00 & 0.98 & 0.99 & 0.99 & 0.99 & 0.94 \\
		
		\hline
		\multirow{9}{1pt}{\rotatebox[origin=c]{90}{Laplace}}&$g_5$& 0 & 0.05 & 0.05 &0.04 &0.05 &0.05 & 0.04& 0.05&0.04 & 0.05 & 0.03 \\
		& $g_5$ & 0.25 & 0.29 & 0.20 & 0.25& 0.28&0.24 &0.11 &0.24&0.16 &  0.20 & 0.14 \\
		&  $g_5$ & 0.5 & 0.80 & 0.58 &0.69 &0.77 &0.69 &0.43 &0.71 &0.51&  0.57 & 0.49 \\
		&  $g_5$ & 0.75 & 0.98 & 0.98 &0.95 &0.97 &0.94 &0.81 &0.96&0.84 &  0.86 & 0.81 \\
		& $g_6(1)$ & 0.1  & 0.15 & 0.15 &0.26& 0.29& 0.27&0.35 &0.39 &0.41 & 0.14 & 0.13 \\
		&  $g_6(1)$ & 0.25& 0.29 & 0.29 &0.38 & 0.41& 0.37&0.38 &0.48 &0.49 & 0.28 & 0.27 \\
		& $g_7$ & 0.5 & 0.68 & 0.68 & 0.88 & 0.85 & 0.87 & 0.70 & 0.79 & 0.78 & 0.74 & 0.60 \\
		& $g_7$ & 0.75 & 0.90 & 0.91 & 0.98 & 0.97 & 0.98 & 0.91 & 0.95 & 0.95 & 0.93 & 0.85 \\
		& $g_7$ & 1 & 0.97 & 0.98 & 1.00 & 1.00 & 1.00 & 0.98 & 0.99 & 0.99 & 0.99 & 0.94 \\
		\hline
		\multirow{9}{1pt}{\rotatebox[origin=c]{90}{Logistic}}&$g_5$& 0 & 0.05 & 0.05 & 0.05 & 0.05 & 0.05 & 0.05 & 0.04 & 0.05 &  0.05 & 0.03 \\
		
		& $g_5$ & 0.25  & 0.14 & 0.13 & 0.16 & 0.18 & 0.16 & 0.07 & 0.15 & 0.13 & 0.15 & 0.10 \\
		&  $g_5$ & 0.5 & 0.46 & 0.41 & 0.47 & 0.51 & 0.48 & 0.27 & 0.45 & 0.37 & 0.46 & 0.35 \\
		&  $g_5$ & 0.75 & 0.77 & 0.76 & 0.80 & 0.84 & 0.80 & 0.59 & 0.78 & 0.68 & 0.79 & 0.68 \\
		
		& $g_6(1)$ & 0.1  & 0.19 & 0.18 &0.07 &0.06 &0.10 &0.04 &0.06 &0.06 & 0.14 & 0.12 \\
		&  $g_6(1)$ & 0.25& 0.27 & 0.27 &0.18 &0.15 &0.17 &0.09 &0.13 &0.13 & 0.27 & 0.25 \\
		
		& $g_7$ & 0.5 & 0.63 & 0.62 & 0.81 & 0.76 & 0.79 & 0.60 & 0.70 & 0.70 & 0.67 & 0.52 \\
		& $g_7$ & 0.75 &0.89  & 0.90 & 0.97 & 0.96 & 0.97 & 0.89 & 0.94 & 0.93 & 0.91 & 0.82 \\
		& $g_7$ & 1 & 0.96 &0.98 & 1.00 & 1.00 & 1.00 & 0.98 & 0.99 & 0.99 & 0.99 & 0.95 \\
		
		\hline
		\multirow{9}{1pt}{\rotatebox[origin=c]{90}{Cauchy}}&$g_5$& 0 &  0.05 & 0.05 &0.04 &0.05 &0.05 &0.04 &0.04 &0.04 & 0.05 & 0.03 \\
		
		& $g_5$ & 0.25 &  0.22 & 0.20  &0.10 &0.14 &0.11 &0.05 &0.12 &0.08 & 0.19 & 0.15 \\
		&  $g_5$ & 0.5  & 0.62 &0.58  &0.28 &0.38 &0.28 &0.14 &0.36 &0.17 & 0.57 & 0.50 \\
		&  $g_5$ & 0.75 & 0.89 & 0.87 &0.50 &0.63 & 0.48& 0.30 &0.64&0.32 & 0.86 & 0.80 \\
		
		& $g_6(1)$ & 0.1  &  0.14 &0.14 &0.06 &0.07 &0.06 & 0.04& 0.06& 0.05 & 0.14 & 0.13 \\
		&  $g_6(1)$ & 0.25& 0.27 &0.12 &0.13 &0.06 &0.06 & 0.13&0.11 &0.08 & 0.28 & 0.26 \\
		
		& $g_7$ & 0.5 & 0.46 & 0.45 &0.59 &0.60 &0.57 &0.37 &0.55 &0.44 & 0.60 & 0.49 \\
		& $g_7$ & 0.75 & 0.79 & 0.77 &0.87 &0.89 &0.86 &0.68 &0.85 &0.74 & 0.89 & 0.80 \\
		& $g_7$ & 1 & 0.93 & 0.92 &0.97 &0.98 &0.97 &0.88 &0.96 &0.90 & 0.98 & 0.94 \\
		
		\hline

		\hline
		
	\end{tabular}
	\label{fig:pow50}
\end{table}

\section{Conclusion}

In this paper we presented two new tests of symmetry based on a characterization and examined their asymptotic properties. 
We calculated the local approximate Bahadur efficiencies of our tests, and performed a small-scale power study.
We found out that our tests are comparable to some commonly used classical tests of symmetry.

When exploring the asymptotics of our tests, the most challenging problem was to obtain the maximal eigenvalue of some integral operators.
In some cases we were able to do it theoretically, using Fourier analysis and a 
decomposition  of linear operators.  For the rest of the cases, we suggested an approximation method based on a discretization of the corresponding 
integral operators. The described procedure could be useful in general for approximating the asymptotic distribution of degenerate 
U-statistics, which emerge often in the problems of goodness-of-fit testing.

\section*{Appendix}

\textbf{Proof of Lemma {\ref{lemaBozin}}.}




To prove the lemma, we show that the local maximum, that coincides with the global one (see Figure \ref{fig: maxEigen}),  is attained at
$t=2/3$ and find its value.

The idea is to demonstrate that both the right and the left derivative of $\nu_1(t)$ at $t=2/3$ are equal to zero.
For this we need the functional equations that $\nu_1(t)$ satisfies in the neighborhood of $t=2/3$. We need both derivatives since these functional
equations happen to be different on different sides of $t=2/3$.

We start from the eigenfunction equation  
\begin{align}\label{eigenEQ}
	\mathcal{Q}(t)[e(x)]=\nu(t)e(x).
\end{align}
Then, for $t$ close to $2/3$ let us decompose the operator
to some simpler operators whose spectra are obtainable in closed form.

\begin{figure}[h!]
	
	\begin{subfigure}{.5\textwidth}
		\centering
		\includegraphics[scale=0.4]{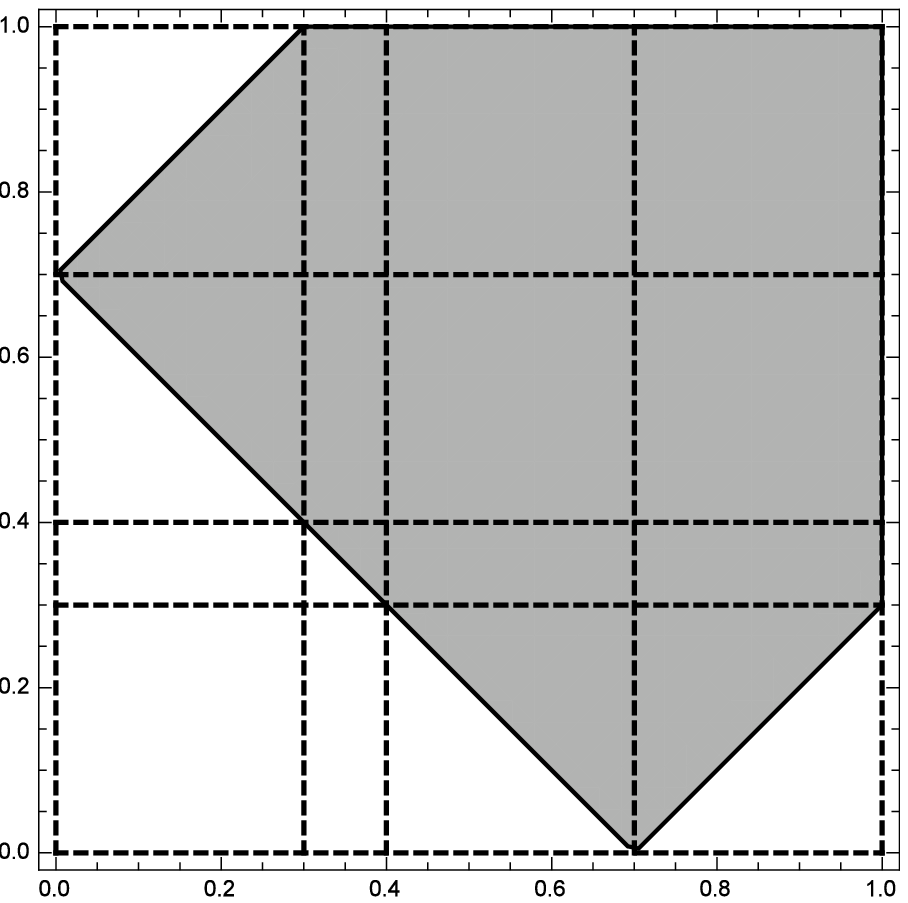}\caption{ The kernel of ${\mathcal{Q}}^{\star}(0.7)$  }
		\label{fig: D70}
	\end{subfigure}
	\begin{subfigure}{.5\textwidth}
		\centering
		\includegraphics[scale=0.4]{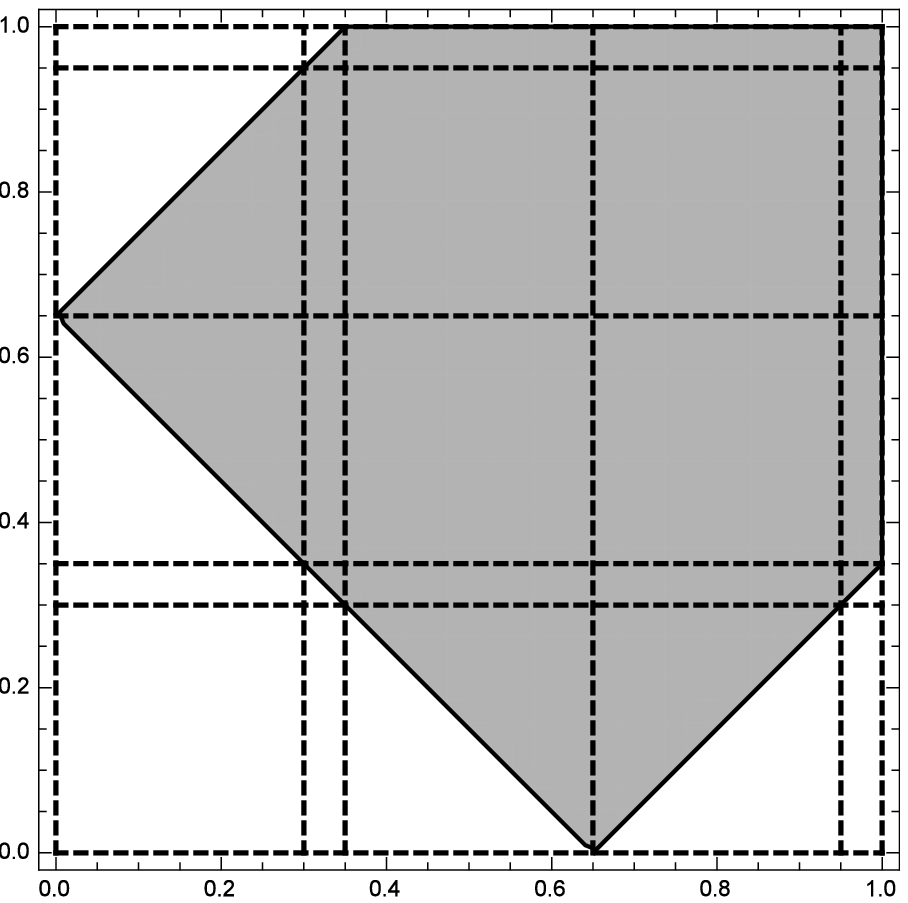}\caption{The kernel of ${\mathcal{Q}}^{\star}(0.65)$  }
		\label{fig: D65}
	\end{subfigure}
	
\end{figure}

First, since the kernels of the family of operators $\mathcal{Q}(t),\;t\in[0,2]$, are odd functions, the corresponding eigenfunctions must be odd, too.
Therefore, instead of $\mathcal{Q}(t)$, we may consider its restriction $\mathcal{Q}^{\star}(t)$, for functions
defined on $[0,1]$, which has the same spectrum. The kernels of the operators $\mathcal{Q}^{\star}(t)$ for $t \in (2/3,1)$ and $t \in (1/2,2/3)$, are shown on 
Figures \ref{fig: D70} and \ref{fig: D65}, respectively. The kernel is equal to one inside the shaded region, and equal to zero outside. 

From Figures \ref{fig: D70} and \ref{fig: D65} one can notice (dashed lines) that the eigenfunctions can be decomposed 
in such a way that the only operators applied to these "subfunctions" are "triangular" or "constant".

We now introduce some notation to formalize this argument. Let $\mathcal{D}_1$ be the "upper right triangular" operator acting on integrable functions $f$
defined on $[0,1]$,
i.e. $\mathcal{D}_1[f](x)=\int_{1-x}^1f(y)dy.$ Analogously, we define the "upper left triangular" operator $\mathcal{L}_1$, and the "lower right  triangular"operator
$\mathcal{R}_1$.
Let $\mathcal{M}_1$ be the mean value operator $\mathcal{M}_1[f]=\int_{0}^1f(y)dy: = \bar{f}$.

We say that two functions $f$ and $g$ defined on $[0,1]$ are {\it reverse} if $f(x)=g(1-x)$, for all $x\in[0,1]$. It is easy to show the following:
\begin{itemize}
	\item[i)] The image  $\mathcal{M}_1[f]$ is a constant function $\bar{f}$;
	\item[ii)] The images  $\mathcal{D}_1[f]$ and $\mathcal{L}_1[f]$ are reverse functions;
	\item[iii)] If the functions $f$ and $g$ are reverse, then $\mathcal{D}_1[f]=\mathcal{R}_1[g]$.
\end{itemize}

Let $f$ be the function defined on $[0,1]$. Denote with $\hat{f}_{a,\delta}$ its contraction to the interval of length $\delta$,  i.e.
for any subinterval
$[a,a+\delta]\subset[0,1]$,
\begin{equation}\label{hatf}
	\hat{f}_{a,\delta}(x):=f((x-a)/\delta).
\end{equation}
Let $\mathcal{D}_{\delta},\mathcal{R}_{\delta},\mathcal{L}_{\delta}$ and $\mathcal{M}_{\delta}$ be the corresponding natural  
$\delta$-contraction operators 
\begin{equation}
	\label{contraction_operator}
	\begin{aligned}
		\mathcal{D}_{\delta}[\hat{f}_{a,\delta}](x)&=\int_{a+b+\delta-x}^{b+\delta}\hat{f}_{b,\delta}(y)dy,\\
		\mathcal{L}_{\delta}[\hat{f}_{a,\delta}](x)&=\int_{b-a+x}^{b+\delta}\hat{f}_{b,\delta}(y)dy,\\
		\mathcal{R}_{\delta}[\hat{f}_{a,\delta}](x)&=\int_{b}^{b-a+x}\hat{f}_{b,\delta}(y)dy.
	\end{aligned}
\end{equation}
The operator $\mathcal{M}$ is a bit different because it can map any
$\hat{f}_{b,\delta}$ to a function defined on an interval of a different length, say
$\hat{f}_{a,\delta_1}$. However, it is a constant operator, so its restriction is just, regardless
of $a$ and $\delta_1$,
$$
\mathcal{M}_{\delta}[\hat{f}_{a,\delta_1}](x)=\int_{b}^{b+\delta}\hat{f}_{b,\delta}(y)dy.
$$
Change of variable in \eqref{contraction_operator} gives us a useful relation
\begin{equation}\label{smena}
	\mathcal{D}_{\delta}[\hat{f}_{b,\delta}](x)=\delta\int_{1-\frac{x-a}{\delta}}^{1}f(y)dy=\delta\mathcal{D}_1[f]\big(\frac{x-a}{\delta}\big).
\end{equation}
The same holds for the other three operators.

Let $t \in[2/3,1]$ (Figure \ref{fig: D70}). We can decompose our eigenfunction $e$ to four subfunctions $\hat{f},\hat{u},\hat{g}$ and $\hat{h}$, as
follows
\begin{multline}\label{ejednacina}
	e(x)=\hat{f}_{0,1-t}(x){\rm I} \{x\in [0,1-t]\}+\hat{u}_{1-t,3t-2}(x){\rm I} \{x\in (1-t,2t-1]+\\
	\hat{g}_{2t-1,1-t}(x){\rm I} \{x\in (2t-1,t]\}+\hat{h}_{t,1-t}(x){\rm I} \{x\in (t,1]\}.
\end{multline}

Applying $\mathcal{Q}^{\star}(t)$ to $e(x)$, for $x \in [0,1-t]$ we get

\begin{align*}
	\mathcal{Q}^{\star}(t)[\hat{f}_{0,1-t}](x)&=\mathcal{D}_{1-t}[\hat{g}_{0,1-t}](x)+\mathcal{R}_{1-t}[\hat{h}_{0,1-t}](x).
\end{align*}
This is exactly what we can see on Figure \ref{fig: D70} for $x\in(0,1-t)$: the first two operators are zero; then comes an upper right operator
$\mathcal{D}$; and finally
the lower right operator $\mathcal{R}$.
On the other hand,  $e(x)$ restricted to this interval is simply $\hat{f}(x)$. Hence from \eqref{eigenEQ} we get
\begin{equation*}
	\mathcal{D}_{1-t}[\hat{g}](x)+\mathcal{R}_{1-t}[\hat{h}](x)=\nu(t)\hat{f}(x).
\end{equation*}

Dilating all the functions to $[0,1]$ using \eqref{contraction_operator} and \eqref{smena}, we get

\begin{equation*}
	(1-t)\mathcal{D}_{1}[g](x_1)+(1-t)\mathcal{R}_{1}[h](x_1)=\nu(t) f(x_1),
\end{equation*}
where $x_1=\frac{x}{1-t}$.

Putting $x$ through all four intervals,  we  transform the equation  \eqref{eigenEQ}  into the system

\begin{align*}
	&(1-t)\mathcal{D}_{1}[g]+(1-t)\mathcal{R}_{1}[h]=\nu(t)f\\
	&(3t-2)\mathcal{D}_{1}[u]+(1-t)\mathcal{M}_{1}[g]+(1-t)\mathcal{M}_{1}[h]=\nu(t)u\\
	&(1-t)\mathcal{D}_{1}[f]+(3t-2)\mathcal{M}_{1}[u]+(1-t)\mathcal{M}_{1}[g]+(1-t)\mathcal{M}_{1}[h]=\nu(t)g\\
	&(1-t)\mathcal{L}_{1}[f]+(3t-2)\mathcal{M}_{1}[u]+(1-t)\mathcal{M}_{1}[g]+(1-t)\mathcal{M}_{1}[h]=\nu(t)h,
\end{align*}
where, for simplicity, we write the equations in terms of the functions only, omitting their arguments.

From the last two equations, using ii), and the fact that the reverse functions have the same mean value, 
we get that $g$ and $h$ are reverse functions. Then, using i) and iii), we transform the system to

\begin{align*}
	&2(1-t)\mathcal{D}_{1}[g]=\nu(t) f\\
	&(3t-2)\mathcal{D}_{1}[u]+2(1-t)\bar{g}=\nu(t)u\\
	&(1-t)\mathcal{D}_{1}[f]+(3t-2)\bar{u}+2(1-t)\bar{g}=\nu(t)g.\\
\end{align*}
Expressing $f$ from the first equation and rearranging the remaining equations we get
\begin{align*}
	((3t-2)\mathcal{D}_{1}-\nu(t)\mathcal{E})[u]&=-2(1-t)\bar{g}\\
	\Big(\frac{2(1-t)^2}{\nu^2(t)}\mathcal{D}^2_{1}-\mathcal{E}\Big)[g]&=-\Big(\frac{3t-2}{\nu(t)}\bar{u}+\frac{2(1-t)}{\nu(t)}\bar{g}\Big),\\
\end{align*}
where $\mathcal{E}$ is the identity operator acting on functions defined on $[0,1]$.
The constant function $\bar{g}$ (and $\bar{u}$) can be expressed as $\bar{g}=\langle g,v\rangle v$, where $v(x)=1$, $x\in[0,1]$,
and $\langle g,v\rangle=\int_{0}^{1}g(x)v(x)dx$ is the scalar product.

Denote, for brevity, $c_1(t)=(3t-2)/\nu(t)$ and $c_2(t)=2(1-t)^2/\nu^2(t)$.
Define the functions $\Psi_1(c_1(t))=\langle v,(\mathcal{E}-c_1(t)\mathcal{D}_1)^{-1}[v]\rangle$
and $\Psi_2(c_2(t))=\langle v,(\mathcal{E}-c_2(t)\mathcal{D}^2_1)^{-1}[v]\rangle$.
Applying the appropriate inverse operators to the left hand side of both equations in the system, and multiplying scalarly with $v$, the system becomes:
\begin{align*}
	\langle u,v\rangle&=c_1(t)\langle g,v\rangle \Psi_1(c_1(t))\\
	\langle g,v\rangle&=c_1(t)\langle u,v\rangle \Psi_2(c_2(t))+\sqrt{2c_2(t)}\langle g,v\rangle \Psi_2(c_2(t)).
\end{align*}
Then, solving the system we obtain the following equation
\begin{equation}
	\label{fundamental}
	1=\Psi_2(c_2(t))\Big(c_1^2(t)\Psi_1(c_1(t))+\sqrt{2c_2(t)}\Big).
\end{equation}
To find the functions $\Psi_1$ and $\Psi_2$ we need the spectrum of $\mathcal{D}_1$.  We get it from the following proposition.

\bigskip

{\bf Proposition. }{\it Let $\{\mu_n\}$ and $\{e_n\}$, $n \in \mathbb{Z}$ be the sequences of eigenvalues and normalized eigenfunctions of $\mathcal{D}_1$.
	Let $v=\sum_{n}a_ne_n$ be the representation of the function $v(x)=1,\;x\in [0,1]$ in the basis $\{e_n\}$. Then $\mu_n=\frac{2}{(4n+1)\pi}$ and
	$a_n=\frac{2\sqrt{2}}{(4n+1)\pi}.$}

The proof can be done either mimicking the proof of Theorem \ref{spektarIntegralne}, or by reducing the eigenfunction equation to 
the appropriate Sturm-Liouville boundary problem.
\medskip

Using the decomposition of linear operators in the basis of its eigenfunctions, and the Proposition, we obtain
$$
\begin{array}{ll}
&\Psi_1(c)=\sum_{n}\frac{a^2_n}{1-c\mu_n}=\frac{1}{c}\bigg(\cot\Big(\frac{\pi}{4}-\frac{c}{2}\Big)-1\bigg),\\
&\Psi_2(c)=\sum_{n}\frac{a^2_n}{1-c\mu^2_n}=\frac{1}{\sqrt{c}}\tan(\sqrt{c}).
\end{array}
$$
For $t=2/3$ the equation \eqref{fundamental} reduces to
\begin{equation}\label{reducedFundamental}
	1=\sqrt{2}\tan\Big(\frac{\sqrt{2}}{3\nu(\frac{2}{3})}\Big).
\end{equation}
The solution with the largest absolute value is $\nu_1(\frac{2}{3})=\frac{\sqrt{2}}{3}(\arctan\frac{1}{\sqrt{2}})^{-1}$.

Using the Implicit function  theorem,  we get that $\nu_1(t)$ is differentiable along $t$ in the right neighbourhood of $t=2/3$.
Furthermore, the right first derivative of right hand side of equation \eqref{fundamental} at $t=2/3$ is equal to
\begin{multline*}
	-\frac{1}{\nu^{2}_{1}(\frac{2}{3})}\bigg(2\tan^2\Big(\frac{\sqrt{2}}{3\nu_1(\frac23)}\Big)-
	3\sqrt{2}\tan\Big(\frac{\sqrt{2}}{3\nu_1(\frac23)}\Big)+2\bigg)-\\- 2\Big(1+\tan^2\Big(\frac{\sqrt{2}}{3\nu_1(\frac23)}\Big)\Big)\frac{\nu'_1(\frac23)}{3\nu^2_1(\frac23)}=-\frac{\nu'_1(\frac23)}{\nu^2_1(\frac23)}.
\end{multline*}
Since this must be equal to zero, we conclude that the right derivative $\nu'_1(2/3)=0$.
The right second derivative of the right hand side of \eqref{fundamental} gives us that $\nu_1''(2/3)<0$. Hence, $t=2/3$ is a "right maximum".

Using a completely analogous procedure, one can show that it is a "left maximum", too, and therefore a local maximum of $\nu_1(t).$ \hfill{$\Box$}

\bigskip

\section*{Acknowledgements}

We would like to thank the referees for their useful remarks that improved our paper.

The research was supported by MNTRS, Serbia, Grant No. 174012 (first author),
MNTRS, Serbia, Grant No. 174012 (second author), and the SPbGU-DFG grant 6.65.37.2017, 
and RFBR grant 16-01-00258  (third author).

\bibliography{simetrija22}
\bibliographystyle{plain}





\end{document}